\documentclass[pre,twocolumn%
,showpacs%
,preprintnumbers%
,amsmath,amssymb,floatfix]{revtex4}

 \usepackage{colordvi}
\usepackage{here}
\usepackage{graphicx}
\usepackage{dcolumn}
\usepackage{bm}
\usepackage{paralist}

\newcommand{\IGN}[1]{{}}
\newcommand{\JTIT}[1]{#1}

\begin{document}

\preprint{Physical Review E vol. 73, in print (2006)}

\title{The role of inhibitory feedback for information processing in thalamocortical circuits }

\author{J\"org Mayer}
\author{Heinz Georg Schuster}%
\author{Jens Christian Claussen}
\affiliation
{Institut f\"ur Theoretische Physik und Astrophysik, Christian-Albrechts Universit\"at, Olshausenstra\ss e 40, 24098 Kiel, Germany}

\date{January 19, 2006}

\begin{abstract}
The information transfer in the thalamus is blocked dynamically
during sleep, in conjunction with the occurrence of spindle waves.
In order to describe the dynamic mechanisms which control the sensory transfer of information, 
it is necessary to have a qualitative model for the response properties of thalamic neurons. 
As the theoretical understanding of the mechanism remains
incomplete, 
we analyze two modeling approaches for  
a recent experiment by Le Masson {\sl et al}.\ 
on the thalamocortical loop. We use a conductance based model in order to motivate an extension of the Hindmarsh-Rose model, which mimics experimental observations of Le Masson {\sl et al}.
Typically, thalamic neurons posses two different firing modes, depending on their membrane potential. 
At depolarized potentials, the cells fire in a single spike mode and relay synaptic 
inputs in a one-to-one manner to the cortex. 
If the cell gets hyperpolarized, T-type calcium currents generate burst-mode firing which leads to a decrease in the spike transfer. 
In thalamocortical circuits, the cell-membrane gets hyperpolarized by recurrent inhibitory feedback loops.
In the case of reciprocally coupled excitatory and inhibitory neurons, 
inhibitory feedback leads to metastable self-sustained oscillations, which mask the incoming input, 
and thereby reduce the information transfer significantly. 
\end{abstract}

\maketitle

\section{Introduction}
\begin{figure}[htbp]
\centerline{\includegraphics[angle=0,totalheight=4.2cm]{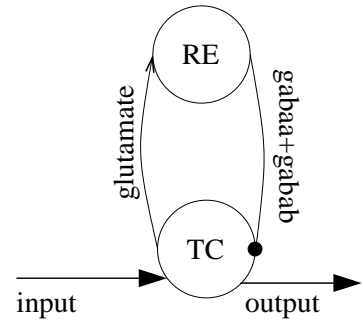}} 
\caption{\label{fig1}Structure of synaptic connections in a pair of reciprocally coupled 
RE-TC cells.}
\end{figure}

\vspace*{-1ex}
Spindle Oscillations are waxing and waning waves that originate in the thalamus at the transition to slow 
wave sleep in mammals \cite{sws}. 
In a pioneering experiment \cite{lem},  
spindle oscillations have  been investigated in a hybrid
thalamic circuit consisting of a biological thalamocortical 
(TC) and an artificial reticular (RE) thalamic cell \cite{lem}. 
An artificial cell was necessary to firstly manipulate selectively 
the synaptic connection of the inhibitory feedback from the RE to the TC cell
(Fig.\ \ref{fig1}).
Spindle oscillations depended critically on the presence of the synaptic connections between 
TC and RE cells \cite{des1,des2}. 
Further, the occurrence of spindle oscillations comes along with a significant 
decorrelation between input and the output of the TC cells. 
Clarification of the relationship between the occurrence of spindle oscillations and 
the decrease in information transmission could help to gain more insight into the mechanisms 
which deprive the sensory information from the consciousness while mammals are sleeping.

Spindle, or bursting, oscillations in excitable media are 
widespreadly observed in a variety 
of physical \cite{amo}, chemical \cite{str}, and biological \cite{sri} oscillators. 
Typically, the potential of a bursting 
neural cell undergoes
subsequent shifting between active and silent phases. 
In the active phase, the membrane potential oscillates quickly, and in the silent phase, 
it evolves slowly without or only with subthreshold oscillations \cite{izh}. 
In this work, we
focus on bursting or spindle oscillations in 
neural systems, 
and investigate their role for information transfer through the
thalamus.
The thalamus, our major gateway for all sensory information,
relays the incoming information depending on the state of arousal \cite{sher,pin}.
To process and relay information \cite{hg1} is a generally important,
but not equally well understood,
feature of neural systems;
here we contribute to the 
understanding of the underlying dynamics of a 
key mechanism in the thalamocortical loop.

The paper is organized as follows. 
In section \ref{sec2}, we use biophysical models of TC and RE neurons interconnected with realistic model synapses as 
introduced in \cite{des1} and also used in the hybrid network investigated by Le Masson in \cite{lem}. 
In section \ref{sec3}, we establish an extension of the Hindmarsh-Rose model \cite{hind}. 
For this purpose, we use dynamically coupled Hindmarsh-Rose neurons to reproduce the 
experimental results of  \cite{lem}. 
We show that the Hindmarsh-Rose system has to be extended by a fourth variable working on a 
very slow timescale, in order to describe the experimental results.

For both models, we investigate the influence of the inhibitory synaptic connection on the information transfer of the TC cell
and compare the results with the experiment of Le Masson {\itshape et al.} \cite{lem}.
An additional
goal is to answer the question whether the complicated biophysical conductance-based model can be replaced by the much simpler 
Hindmarsh-Rose model without loss of the most important dynamical features. 
We show how to achieve this by introducing an extended Hindmarsh-Rose model
with an additional degree of freedom.
Some important features are
better reproduced by the extended Hindmarsh-Rose neuron than by the biophysical
model.
Our model is by far less computationally costly, 
yet simple enough to understand dynamical mechanisms.

\section{Information Transfer in a biophysical model of a thalamocortical oscillator\label{sec2}}
A common approach to model the thalamocortical system
are conductance-based models \cite{rabino,bal1,des3,hod}.
Descriptions of the properties of  single neural cells can be found for example in \cite{des1,des2}.
The case of two interconnected  cells  without external forcing has been investigated in \cite{des2}.
To our knowledge the case of two interconnected cells as shown in 
Fig.\ \ref{fig1} which get excited by an unregularly distributed realistic
synaptic bombardment as used in the experiment of Le Masson {\itshape et al.} \cite{lem} has not been studied in detail up to now. 
In this section we will investigate the influence of the inhibitory synaptic connection on the information transfer of the TC cell for this case.
Destexhe {\itshape et al.} \cite{des2} model the TC cell and the RE thalamic cell by using a 
conductance-based single compartment model. 
The time evolution of the membrane potentials 
is governed by the following cable equations
\begin{eqnarray}
\label{eq1}
C_{m}\dot{V}_{T}&=&-I_{T_{L}}-I_{T}-I_{h}-I_{T_{Na}}-I_{T_{K}}\nonumber\\
                & &-I_{GABA_{a}}-I_{GABA_{b}}\\
C_{m}\dot{V}_{R}&=&-I_{R_{L}}-I_{Ts}-I_{K[Ca]}-I_{CAN}-I_{R_{Na}}-I_{R_{K}}\nonumber\\
                & &-I_{GLU},
\label{eq2}
\end{eqnarray} 
where $V_{T}$ is the membrane potential of the TC cell; $C_{m}$ is the capacity of the membrane. According to \cite{des2}, 
we included the following ion currents: T\begin{scriptsize}\end{scriptsize}he leakage current $I_{T_{L}}$ and $I_{R_{L}}$, the low-threshold 
$Ca^{2+}$ currents $I_{T}$ and $I_{Ts}$, the hyperpolarization activated current $I_{h}$, 
the $Ca^{2+}$-activated currents $I_{K[Ca]}$ and $I_{CAN}$, and, like in the Traub and Miles \cite{tra} model, 
the fast $Na^{+}$ and $K^{+}$ currents $I_{T/R_{Na}}$ and $I_{T/R_{K}}$, which are responsible for the generation of action potentials.
The synaptic currents $I_{GABA_{a}}$ and $I_{GABA_{b}}$  represent the $GABA_{a}$ and $GABA_{b}$ receptors 
in the synapses from RE to TC cells, while $I_{GLU}$ describes the excitatory synapse from the TC to RE cells.
 
Destexhe {\itshape et al.} \cite{des2} refer to several sources for the description of the ion currents. The kinetic equations of the TC cell have been described in detail in \cite{des3}.  The $I_{T}$ kinetics was taken from Wang {\itshape et al.} \cite{wan1}. $I_{h}$ was described by the model of Destexhe {\itshape et al.} \cite{des3}, which incorporates a regulation by intracellular calcium.
For the RE cell $I_{Ts}$ was taken from Huguenard and Prince \cite{hug1}. The kinetics of the $Ca^{2+}$-dependent currents $I_{K[Ca]}$ and $I_{CAN}$ are adjusted to the clamp data of RE neurons \cite{bal2}. For both cells, $I_{Na}$ and $I_{K}$ were taken from \cite{tra}. The calcium dynamics in both cells are described by a simple model which was introduced in \cite{des3}.
All details are described in appendix \ref{secA}.

In biology, the typical input signal for the TC cells is a spike train with unregularly distributed inter-spike intervals, which can be modeled by a Poisson process with a refractory period. 
For example, the retinal cells fire such unregularly distributed spikes in darkness or under constant illumination. 
The refractory period is necessary because every biological cell needs some time $\tau_{r}$ to recover after it has fired a spike, 
so the cell cannot fire spikes with arbitrary low inter spike intervals. 
In our model, the ISI are distributed according to the following distribution
\begin{equation}
\tilde{f}(\tau)=\begin{cases}
0   &
\textnormal{for } \tau< \tau_{r}\\
\bar{r} e^{\bar{r}  \tau_{r}} e^{-\bar{r}  \tau} &
\textnormal{for } \tau \geq \tau_{r}
\end{cases},
\end{equation}
where $e^{\bar{r} \tau_{r}}$ is a scaling factor, 
 $\bar{r}=1/100$ and $\tau_{r}=30ms$ here. 
If we stimulate our model by such a modified Poisson process, we have a computational model for the experiment of Le Masson {\itshape et al.} \cite{lem}, which allows us to compare our computational model with the experiment.

\subsection{Simulation results of the biophysical model:
Spindle oscillations, waxing and waning, 
and influence on information transfer}
With the  TC and RE cells interconnected, the computational thalamic circuit receiving this realistic synaptic 
bombardment 
\cite{lem}
showed 
waxing and waning oscillations. 
These spindle sequences consisted of bursts of 8-10Hz oscillations lasting a few seconds, and were separated by silent phases of around 8-10s. 
The periodical occurrence of this oscillations is very similar to biological spindle waves which occur in sleep-related states. 
As in the experiment, each spindling state is terminated by an after-depolarization of the membrane potential. 
The reason for this effect is the hyperpolarization activated current $I_{h}$ 
(see appendix \ref{secA}); when this after-depolarization
adds to the synaptic bombardment, it forms a depolarization which deactivates 
the low threshold $Ca^{2+}$ current $I_{T}$ and subsequently blocks the input signal transfer.
\begin{figure}[htbp]
\begin{center}
\includegraphics[totalheight=10.5cm]{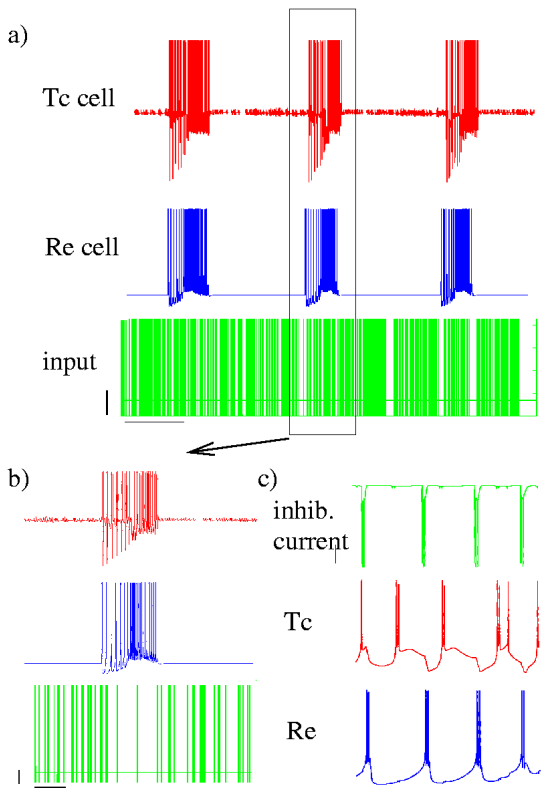}
\caption{\label{fig2}(Color online) {(a)} Spontaneous spindle activity in the computational thalamic circuit. 
The TC cell receives artificial synaptic retinal bombardment, modeled by a Poisson-distributed 
spike train.  The RE-TC oscillator shows spindle activity. {(b)} Detail of (a).
{(c)} Total synaptic current injected into the biological 
TC cell and simultaneous voltage of the TC and RE cell. 
Bottom: The hyperpolarization of the TC cell activates the $I_{T}$ current, what initiates a postinhibitory rebound burst 
(see also Fig.\ \protect\ref{figschema}). Calibration bars: 10s, 20mV (a); 1s, 20mV (b);  10ms, 6nA, 20mV, 1nA (c).} 
\end{center}
\end{figure}
\subsection{Measuring the information transfer}
During the spindling state, the firing pattern of the TC cells (excitor), which is very different from the input 
(see Fig.\ \ref{fig2}a--b),
shows that obviously
the information transfer of the input is low. Le Masson {\itshape et al.} \cite{lem} answer
the question whether this low transfer is still reliable, by calculating two different indices. 
First, the contribution index ($T_{CI}$) examines the TC cell discharge and quantifies the percentage
of output spikes $N_{out}$ which are precisely correlated with retinal input spikes. It estimates the reliability with which a TC spike can be considered as being triggered by an input spike rather than being spontaneous. It is computed as the peak of the crosscorrelation, normalized by the number of output spikes. Second, the correlation index ($T_{CC}$) measures the global efficiency of the input-output spike transfer and indicates the ratio of input spikes being actually transmitted as output spikes in the TC neuron. It is computed as the peak of the crosscorrelation between retinal and TC neuron spikes, normalized by the number of retinal cell spikes. To be able to compare our numerical results with the experimental results of Le Masson {\itshape et al.} \cite{lem} quantitatively, we will use exactly these measures here. 
\begin{figure}[htbp]
\begin{center}
\includegraphics[totalheight=5.5cm]{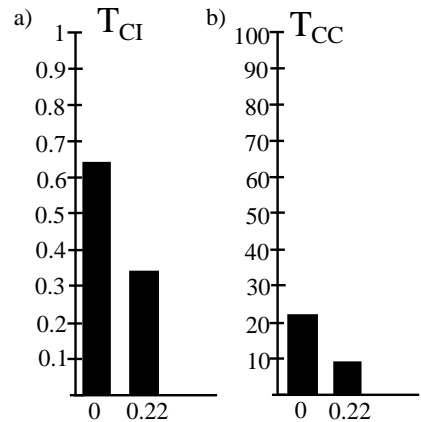}
\caption{\label{cc_ci}(Color online) Averaged contribution ($T_{CI}$) (a) and correlation ($T_{CC}$) (b) indices versus synaptic strength.} 
\end{center}
\end{figure}

To be able to compare the numerical results  obtained from the biophysical model with  the simplified model introduced in this paper, we will define two new measures for the information transfer which are equivalent to the measures used by Le Masson {\itshape et al.} in \cite{lem}, but do not depend on the detailed form of the signals.
First, the signal-to-noise ratio $(T_{SN})$ examines the amplitude of the TC-cell signals and quantifies the percentage
$N_{tr}$ of 
output spikes $N_{out}$ which are exactly triggered
by an input spike; such it is a measure equivalent to the $T_{CI}$ used. 
An output spike is considered to be triggered by an input spike if it occurs within a delay of $<50ms$ after an input spike. 
As the $T_{CC}$, it is a measure for the reliability with which an output spike can be considered as being triggered by an input spike rather 
than being due to the intrinsic dynamic of the circuit.
\begin{equation}
{\rm T_{SN}}=\frac{N_{tr}}{N_{out}}.
\end{equation}
Second, the transfer efficiency $T_{TE}$ measures the global efficiency in the input-output spike transfer, and 
indicates the ratio of input spikes  being actually transmitted as output spikes, obviously it is equivalent to the measure $T_{CC}$. It is computed by the number of triggered spikes $N_{tr}$ divided by the total number of input spikes,
\begin{equation}
{\rm T_{TE}}=\frac{N_{tr}}{N_{in}}.
\end{equation}
The difference between the measures $(T_{SN})$ and ($T_{CI}$) is that our measure checks if an output spike occurs within a certain delay after an input spike, while the contribution index counts the spikes at a fixed time delay defined by the maximum of the crosscorrelation. The $T_{CC}$ and $T_{TE}$ differ in the same way as mentioned above, further $T_{TE}$ only takes values between zero and one. As our measures only count the spikes, and such treat spikes like binary events, they can be used for any spiking system and allow a comparison of the transmission properties of different models. The disadvantage is that they do not shed light on the amplified details of a signal system such as $T_{CI}$ and $T_{CC}$ do. 

\begin{figure}[htbp]
\begin{center}
\includegraphics[angle=0,totalheight=6cm]{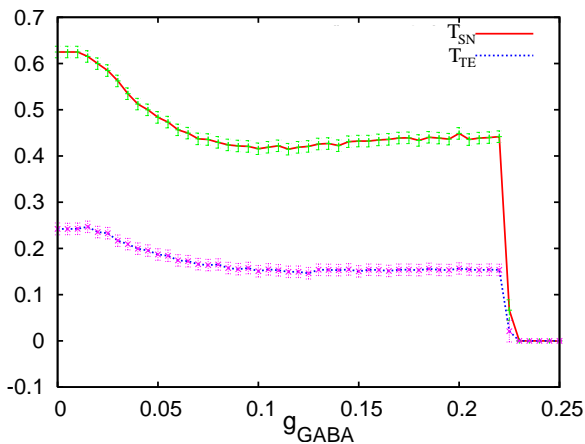}
\caption{\label{figSNTEbio}(Color online) Percentage of output spikes triggered by an input spike $(T_{SN})$ 
(solid) and Transfer Efficiency $T_{TE}$ 
(dashed) for different values of the inhibitory feedback. 
Both indices show no significant change when the inhibitory feedback is varied within the stability borders.}
\end{center}
\end{figure}

These two measures, the  transfer efficiency and the transfer reliability, are necessary to characterize the information transfer appropriately. The usage of two different transfer measures allows to answer the question whether the transfer gets blocked or if incoming signals get masqued by autonomous oscillations of the system. 
The results are given in Fig.\ \ref{figSNTEbio}.

In the presence of strong inhibitory feedback ($g_{GABA_{a}}=0.1mS$ and $g_{GABA_{b}}=0.01mS$) the ($T_{CI}$) and $(T_{SN})$ were 
low ($\approx0.4$), indicating that less than half of the output
spikes were triggered by an input spike; thus the thalamus is not transferring spikes in a one-to-one manner. 
To answer the question whether the degree of inhibition produced by the RE neurons could directly control 
the precision of spike transfer, we varied the strength of inhibition. Our simulations show that the 
influence of the inhibitory feedback is 
by far not so strong as in the experiment
\cite{lem}. In our computational model, the $T_{SN}$ varied only between $\approx0.6$ and $\approx0.4$. 
In agreement with the experimental data
\cite{lem},
the global efficiency of spike transfer $T_{TE}$  was not significantly different in the absence or presence of
inhibitory feedback.

\subsection{Discussion of the biophysical model} 
Altogether, our simulation cannot reproduce the transmission behavior of the thalamic 
circuit observed
\cite{lem}
experimentally. When  we compare the numerical achieved values for the $T_{CI}$ and $T_{CC}$ (Fig.\ \ref{cc_ci}) quantitatively with the values from the experiment, we find that the effect of input-output 
spike decorrelation is present, but lacks some features. First, even without or very low
inhibitory feedback, the $T_{CC}$ is 
not more than $\approx 0.25$; in the experiment the $T_{CC}$ varied in  a range from about 85--45 (see Figure 3e in \cite{lem}). For a biological system this would mean that the animal would 
always be drowsy. Second, the system does not allow precise up and down regulation of the 
signal reliability in a wide range, the $T_{CI}$ only varies in range from around 0.7--0.35, that is 
about a factor 2, while in the experiment the $T_{CI}$ varies from about 0.8--0.2 (see Figure 3d in \cite{lem}) which 
corresponds to a factor 4. Further, spindle oscillations occurred even with very little 
or no inhibitory feedback, which is also in contrast to the experimental results.

\begin{figure}[htbp]
\begin{center}
\includegraphics[angle=270,totalheight=6cm]{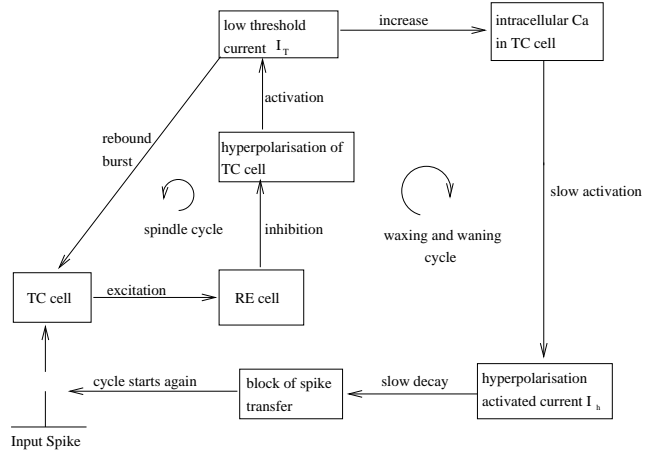}
\caption{\label{figschema}Schematic diagram illustrating the generation mechanism of spindle oscillations. We distinguish two cycles: 
First, the spindle cycle is responsible for the spiking; the system runs several time through it before the waxing and waning cycle is activated, leading to bursts of oscillations. The slowly decaying current $I_{h}$ leads to the long silent epochs during which spike transfer is blocked. 
Altogether, the interplay of the currents $I_{T}$ and $I_{h}$ plays an important role for the genesis of spindle oscillations.
 (see also Figs.\ \protect\ref{fig2} and \protect\ref{figSNTEbio}).}
\end{center}
\end{figure}

At this point it appears questionable whether the
detailed biophysical model described above
---~with its high dimensionality and parameter space
quite too detailed to gain systematic understanding~---
provides the appropriate level of description to
analyze coupled systems, e.g.\ of the
thalamocortical loop investigated here.
We will however use the biophysical model to 
establish our simplified model.
As the biophysical model incorporates many features of realistic biological neurons, it allows us to identify 
the key elements which may be responsible for sleep spindle oscillations. The slow repetition rate of 
spindling (0.3--0.1 Hz) is due to intrinsic mechanisms of the TC cell. 
The slow variables $F_{2}$ and $S_{2}$ 
(Fig.\ \ref{fig20})
of the $I_{h}$ current 
(see \ref{ext} and appendix A) play the key role for these long silent phases. 
The spindling starts in the TC cell due to the hyperpolarization low threshold $Ca^{2+}$ current $I_{T}$ 
(see Fig.\ \ref{fig2}c and Fig.\ \ref{figschema}). The spindling terminates due to the activation of the $Ca^{2+}$ activated current $I_{H}$ 
(see Fig.\ \ref{figschema}). The frequency of the spindling is determined by intrinsic properties of the RE cell, as shown by Destexhe 
in \cite{des2}, but also the intrinsic properties of the inhibitory synapses play a role. 
To provide an overview, Fig.\ \ref{figschema} shows a schematic diagram of the mechanisms leading to spindle oscillations.

Unfortunately, 
due to the complexity of the
high-dimensional biophysical system, the connection between biophysics and dynamics 
is hard to unravel. 
Further, the computational effort is very high,
making studies of networks almost impossible. 
To solve these problems, we
propose a reduced model
to reproduce the experimental data of 
Le Masson {\itshape et al.} by using the
simpler Hindmarsh-Rose \cite{hind} type neurons, and modify them 
in a way that the 
features described above are reproduced.

\section{The reduced system: An extended Hindmarsh-Rose model\label{sec3}} 

\subsection{The Hindmarsh-Rose model\label{sec31}}
The Hindmarsh-Rose equations are a simple polynomial model of bursting in thalamic cells, which reproduce several features, 
like for example rebound bursts, of more complicated biophysical models \cite{wil,des1,dham}.  
The original Hindmarsh-Rose equations are given by
\begin{eqnarray}
\label{eq7}
\dot{v}&=&w-v^3+3v^2-z+I(t)\\
\label{eq8}
\dot{w}&=&1.8-5v^2-w\\
\label{eq9}
\dot{z}&=&\epsilon(3.3(v+1.56)-z), 
\end{eqnarray}
where $v$ is the voltage, $w$ is a recovery variable, $I(t)$ is an external forcing and $z$ is a slow variable. 
In order to understand the behavior of equations (\ref{eq7}--\ref{eq9}), we first consider the limit $\epsilon\longrightarrow 0$, 
so $z(t)=z$, which results in a planar reaction-diffusion system. This fast subsystem is responsible for the spiking 
dynamics of the Hindmarsh-Rose system. The dynamics of this subsystem is also crucial for $\epsilon\neq0$. This procedure is called singular approximation, or slaving principle, and is widely used to separate fast and slow subsystems \cite{pon,zee,hak}. Slow variables (here $z$) can be treated as slowly varying parameters and the rest of the system can be studied as a function of these new parameters. We apply this method to the Hindmarsh-Rose model.

To understand the influence of the variable $z$ on the $(v,w)$ subsystem, 
equations (\ref{eq7}--\ref{eq8}) with $I(t)=0$ are
transformed to a Lienard system \cite{sri,dipl}, yielding
(App.\ \ref{secB2})
\begin{eqnarray}
0=\ddot{v}+f(v)\dot{v}+g(v),
\label{eq11}
\end{eqnarray}
where $g(v)=v^{3}+2v^{2}-1$ can be considered as the gradient of a potential $\Phi(v,z)$, 
and $f(v)=1-6v+3v^{2}$ as a damping term. The roots of $g(v,z)$ give the fixed points of the equation. 
Depending on the parameter $z$, there exist either one or three real roots.

\begin{figure}[htbp]
\begin{center}
\includegraphics[angle=0,totalheight=6cm]{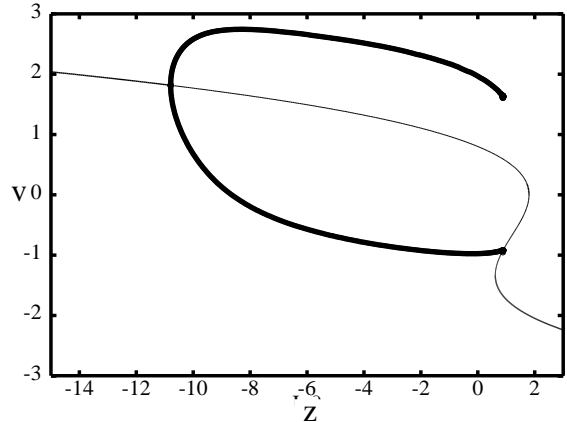}
\caption{\label{fig6}Maximum/Minimum bifurcation diagram of $v$ in the fast subsystem (4) as a function of $z$. Heavy lines indicate stable fixed points (limit cycles) and thin lines indicate unstable fixed points (limit cycles).
(See Appendix \ref{secB3}.)}
\end{center}
\end{figure}

The bifurcation diagram of 
equations (\ref{eq7}--\ref{eq8}) as a function of $z$ is shown in 
Fig.\ \ref{fig6} (the bifurcation diagram was computed with the auto interface \cite{auto} of xpp \cite{xpp}). 
If we start with $z=2$ from the fixed point $(v=-2.05,\;w=-19.67)$ of the fast subsystem and go to the left in the bifurcation diagram, 
the stable fixed point loses its stability by a saddle node bifurcation at $z\approx0.6148$. 
For values of $z$ between $[0.6148,0.8891]$, a stable limit cycle and a stable fixed point coexist. 
At $z=-10.4$, the upper branch of the steady state loses stability by a Hopf bifurcation (for further details see \cite{wang,sri}). Here we will only focus on values of $z$ between $[-5,+5]$.\\

\subsection{Periodic forcing of the Hindmarsh-Rose neuron} 
Another important case 
is that of $z$ being a periodic function $p(t)$; 
this is the case for certain values of a constant external forcing $I$ \cite{wang}, or when the neuron is coupled to another periodic spiking or bursting neuron. So in a next step we consider the fast $(v,w)$-subsystem with periodic forcing
\begin{eqnarray}
\dot{v}(t)&=&w-v^3+3v^2+p(t)\nonumber\\
\dot{w}(t)&=&1.8-5v^2-w,
\label{eq12}
\end{eqnarray}
or, in Lienard form
\begin{equation}
p(t)=\ddot{v}+f(v)\dot{v}+g(v),
\label{eq13}
\end{equation}
where $f$ and $g$ are defined as above. According to \cite{far}, theorem 4.3.1 and 4.3.3,  
equation (\ref{eq13}) has a non-constant periodic solution with the same period as the forcing term $p(t)$ if several conditions  (i-vi) (given in Appendix \ref{secB4})
are fulfilled. 
Hence, we make the following proposition: When $p(t)$ is a $T$-periodic function, 
then (\ref{eq13}) has a periodic solution with the same period $T$.
The detailed proof is in Appendix \ref{secB4}.

\begin{figure}
\begin{center}
\includegraphics[angle=0,totalheight=6cm]{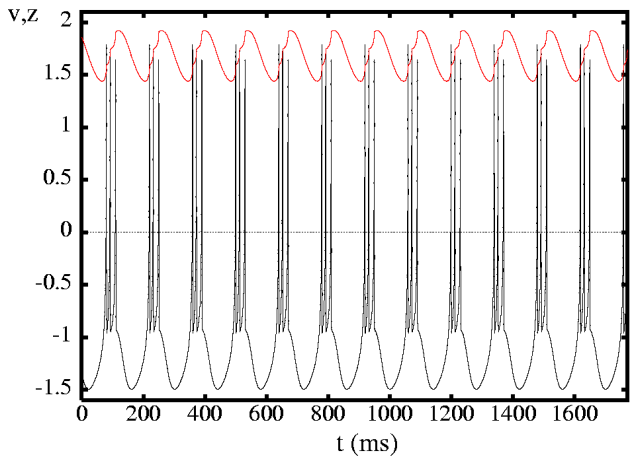}
\caption{\label{fig7}(Color online) $v(t)$ 
(black) and $z(t)$ 
(red/gray) 
with $I=1.0$. When $z(t)$ is periodic, $v(t)$ oscillates with the same period; thus the slow variable $z(t)$ slaves $v(t)$.}
\end{center}
\end{figure}

For illustration, we show the behavior of the Hindmarsh-Rose equations (\ref{eq7}--\ref{eq9}) excited with a constant forcing $I(t)=I_{0}$
(see Fig.\ \ref{fig7}),
thus again $z(t)$ is a periodic function. 
This is an example for a typical intrinsic burster, where the slow variable slaves the fast subsystem \cite{wang}.
The theorem presented above proves that the fast $(v,w)$-subsystem of the Hindmarsh-Rose model shows mode locking when it is driven by a periodic forcing, no matter what frequency or amplitude this forcing has.

If we assume  $\epsilon$ to be small ($\approx0.006$ as used by \cite{ros}), $z$ can be treated like a slowly varying parameter 
in the $(v,w)$ system (\ref{eq12}). To understand the influence of $v$ on 
equation (\ref{eq9}), we write it in the form of a relaxation process 
\begin{equation}
\dot{z}=-\epsilon(z-z_{\infty}(v)),
\label{eq14}
\end{equation} 
where $z_{\infty}(v)=3.3(v+1.56)$. 
As $\epsilon$ is very small, the fast dynamics of $v$ has only little influence on 
equation (\ref{eq14}), as its own dynamics is too slow to follow the fast $z_{\infty}(v)$. 
Because of this, $z$ will arrive in a steady state if $v$ is spiking quickly, 
and then just oscillate 
with a very little amplitude around it. This effect leads to the so-called spike frequency adaption 
(for details see \cite{hind}), which is also observed in biological neurons \cite{wil}. 
For small enough values of $I(t)$ this effect can also lead to an effect similar to intrinsic bursting \cite{wang}. 

\subsection{Post-inhibitory rebound bursts in the Hindmarsh-Rose neuron}
\begin{figure}
\begin{center}
\includegraphics[angle=0,totalheight=6.5cm]{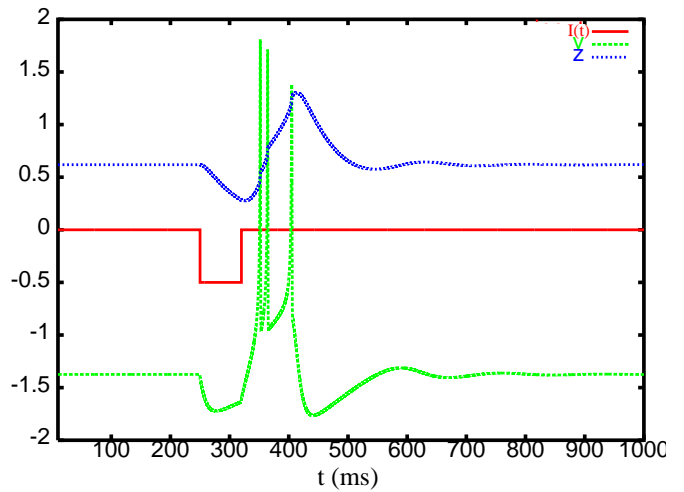}
\caption{\label{fig8}(Color online) $v(t)$, $z(t)$ and  $I(t)$. 
The system shows a post-inhibitory rebound burst after a hyperpolarizing step of $I(t)=-0.5$ and a duration of $70ms$.}
\end{center}
\end{figure}

\begin{figure}[htbp]
\begin{center}
\includegraphics[angle=0,totalheight=4.5cm]{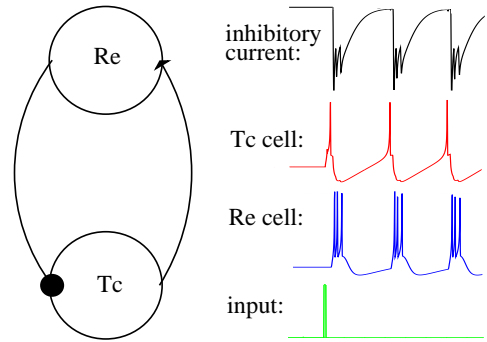}
\caption{\label{fig9}(Color online) Right: Reciprocally coupled RE and TC cell. 
Left: If the excitor cell gets activated by an external pulse, 
it excites the inhibitor cell, which leads to an inhibitory current. This current hyperpolarizes the excitor cell and evokes a rebound burst in it, and the mechanism repeats, which leads to 
self-sustained oscillations.} 
\end{center}
\end{figure}

Post-inhibitory rebound bursts are a dynamical
feature being 
also present in the biophysical neuron model \cite{des1}.
As it is crucial for spindle oscillations and for the information transfer, 
we will describe it in detail here, see also Fig.\ \ref{fig8}. 
If the Hindmarsh-Rose model \cite{hind} is hyperpolarized 
for a period similar to the burst duration, 
the adaption current $z$ will decrease below its steady state. If the hyperpolarizing current is released, the $z$ current 
stays below its steady state for some time due to its slow timescale.
As a consequence, the model behaves as if some extra current had been applied, so it will cross the saddle node bifurcation in 
Fig.\ \ref{fig6} and stay on the limit cycle. 
Due to the spikes, $z$ will increase above its steady state, stopping the burst 
(see Figs.\ \ref{fig8},\ref{fig9}). 
Due to the slow timescale, a postinhibitory rebound burst will only appear if 
slowly-decaying dynamic synapses are used, as otherwise $z$ will not decrease below its steady state. 
In the biophysical model, the same effect occurs due to the low threshold calcium current $I_{T}$ 
(see Fig.\ \ref{fig2}c) \cite{des1}, so the variable $z$ can be interpreted as a calcium current.

\begin{figure}[htbp]
\begin{center}
\includegraphics[angle=0,totalheight=12cm]{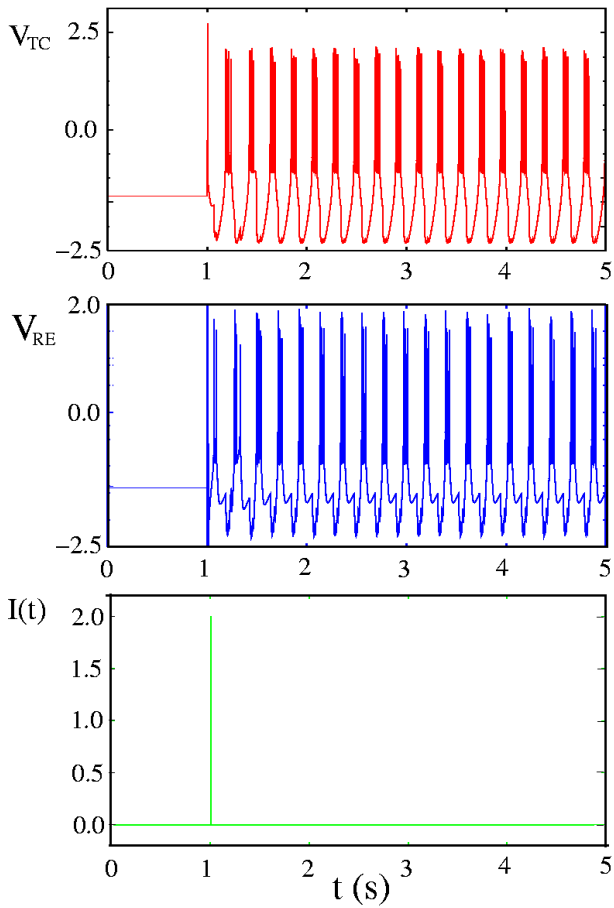}
\caption{\label{fig10}(Color online) Reciprocally coupled inhibitor and excitor; without the $h$ current, the oscillation does not terminate.
}
\end{center}
\end{figure}

In reciprocally coupled neurons, the effect described 
above leads to self-sustained oscillations if the excitor gets activated by a single input spike; this effect is an example for 
so-called hard-excited self-oscillations \cite{hak}. During this oscillation, the two neurons 
exhibit anti-phase synchronization. We show this for the case of two Hindmarsh-Rose neurons, which are coupled like in a typical thalamocortical circuit (see Fig.\ \ref{fig9}).

\subsection{Motivation of a simplified calcium current: Extending the Hindmarsh-Rose model}\label{ext}
\begin{figure}
\begin{center}
\includegraphics[angle=0,totalheight=12cm]{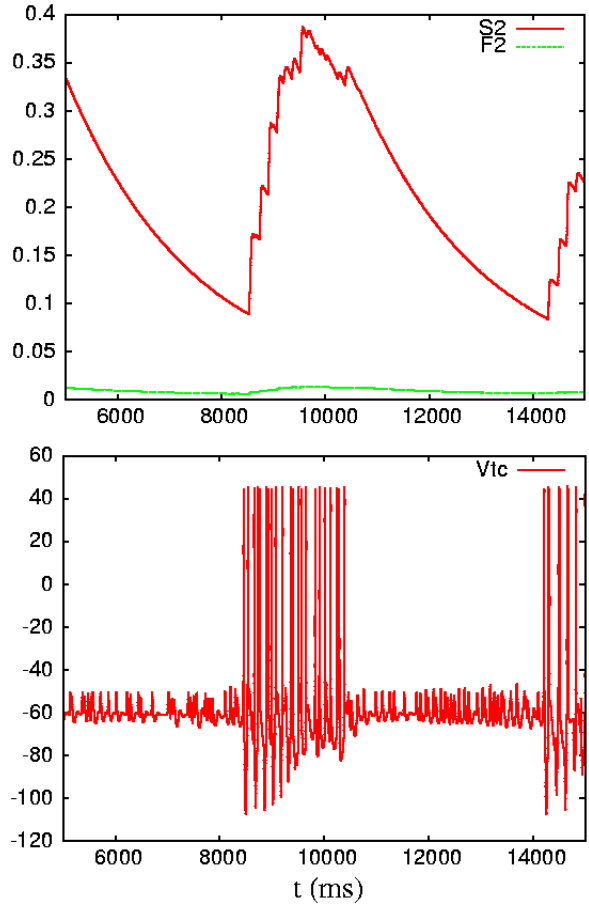}
\caption{\label{fig20}(Color online) Top: Time course of the gating variables $S_{2}$ (red) and $F_{2}$ (green) during a spindle cycle. Bottom: The membrane potential $V_{Tc}$  of the biophysical model}
\end{center}
\end{figure}

The self-sustained oscillations in 
Fig.\ \ref{fig10} are stable and do not terminate; this is contrary to the experimental observations of Le Masson {\itshape et al.} \cite{lem}. 
A spindle oscillation or a burst consists of a series of metastable self-oscillations enhanced by rebound bursts which terminate after  a few seconds. 
Above we argumented that the slow variable $z$ can be interpreted as a calcium current, and such corresponds to the low threshold calcium current $I_{T}$ in the biophysical model. The schematic diagram Fig.\ \ref{figschema} shows that the reason for the termination of the oscillation is a further ion current $I_{h}$ \cite{lem,des1,des2,ahl}. As the original Hindmarsh-Rose model does not incorporate a variable which corresponds to the $I_{h}$ current in the biophysical model, we will extend the Hindmarsh-Rose model by an equation motivated directly by the dynamics of $I_{h}$ in the biophysical model. Destexhe {\itshape et al.} \cite{des2} model $I_{h}$ by a double activation kinetic, consisting of slow and fast activation variables, regulated by intracellular $Ca^{2+}$,
\begin{eqnarray}
S_{0}\leftrightarrows S_{1}\\
F_{0}\leftrightarrows F_{1},
\end{eqnarray}
were $S_{0/1}$ and $F_{0/1}$ represent the closed and open states of the slow and fast activation gates, respectively. The open state gates are assumed to have $n$ bindings for $Ca^{2+}$ which lead to the open bounded gates $S_{2}$ and $F_{2}$,
\begin{eqnarray}
S_{1}+nCa^{2+}\leftrightarrows S_{2}\\
F_{1}+nCa^{2+}\leftrightarrows F_{2}.
\end{eqnarray}
In this model the activation function of $I_{h}$ is shifted during the oscillatory phase by the entry of $Ca^{2+}$, and thus terminates the oscillation. Destexhe {\itshape et al.} \cite{des2} find that the length of the silent phase and of the oscillatory phase were directly proportional to the time constant of intracellular $Ca^{2+}$ binding to $I_{h}$ channels, $k_{2}^{-1}$. Further it is assumed that the binding of $Ca^{2+}$ is critical for the onset and termination of the oscillatory phase \cite{des2}. So the length of the oscillatory phase depends on the rate of rise of the variables $S_{2}$ and $F_{2}$, whilst the length of the silent phase depends on the rate of relaxation of $S_{2}$ and $F_{2}$ back to their resting values (see Fig.\ \ref{fig20} and \cite{des2}).
According to (\ref{eqA21}), both the length of rising and relaxation of $S_{2}$ (red) and $F_{2}$ are proportional to $k_{2}^{-1}$. As Fig.\ \ref{fig20} shows, $F_{2}$ only displays small variations of amplitude compared to $S_{2}$ and therefore plays a less important rule. In our further simplification we will neglect the influence of $F_{2}$. The slow variable $S_{2}$ slaves the system and switches between the oscillatory and the resting state leading to waxing and wanning oscillations \cite{des2}. Our simplified model will be motivated from the kinetic equation of $S_{2}$.
As the transition from $S_{0}$ to $S_{1}$ is much faster than the one from $S_{1}$ to  $S_{2}$, we will assume it to be instantaneous, what leads to $S_{2} \approx S_{0}$, so we get
\begin{equation}
S_{0}+nCa^{2+}\leftrightarrows S_{2}
\end{equation}
In equation (5) in \cite{des2}, the number of binding sites $n$ for $Ca^{2+}$ is assumed to be $2$, in our model we use $n=1$ with the backward rate  $k_{2}=4*10^{-4}$, the forward rate $k_{1}=\frac{k_{2}}{5*10^{-4}}$ and $C=\frac{[Ca_{in}]}{5*10^{-4}}$, where $[Ca_{in}]$ is the intracellular calcium concentration, thus altogether we get
\begin{equation}
\dot{S_{2}}=-k_{2}\left[ S_{2}-CS_{0} \right] \label{eq_cal1}.
\end{equation}
The essential features of equation (\ref{eq_cal1}) can be summarized as follows:
$S_{2}$ opens and closes proportional to the same rate constant $k_{2}$; due to $C$, the activation depends critically on the concentration of intracellular $Ca^{2+}$. This features should also be present in our simplified model. In the Hindmarsh-Rose model the calcium current is mimicked by the variable $z$ \cite{hind}, in our model for the $h$ current will get activated by the variable $z$. The rate of activation and deactivation will be equal to the constant used in the biophysical model. For convenience and to be consistent with the notation used in biophysics,
we call this current $h$.
\begin{equation}
\dot{h}_{TC}=-k_{2}(h_{TC}-0.88(0.9-z_{TC})).\label{eq1n}
\end{equation}
As $z$ decreases when the cell gets hyperpolarized, we use the difference between the maximum magnitude of $z$ and $z$ itself as an activation term in (\ref{eq1n}). 
In order to reproduce the behavior of  the biological 
TC cell, we  extend the Hindmarsh-Rose equations as follows. 
\begin{eqnarray}
\dot{v}_{TC}&=&w_{Tc}-v_{TC}^3+3v_{TC}^2-z_{TC}-h_{TC}
\label{eq17a}\\
\dot{w}_{TC}&=&1.8-5v_{TC}^2-w_{TC}
\\
\dot{z}_{TC}&=&\epsilon\left[ 3.3(v_{TC}+1.56)-z_{TC}\right]
\\
\dot{h}_{TC}&=&-0.0004(h_{TC}+0.88(0.9-z)). \label{eq17d}
\end{eqnarray}
Equations (\ref{eq17a}-\ref{eq17d}) define our central model,
and will be used to model
the reciprocally coupled neurons in the remainder.
\begin{figure}
\begin{center}
\includegraphics[angle=0,totalheight=12cm]{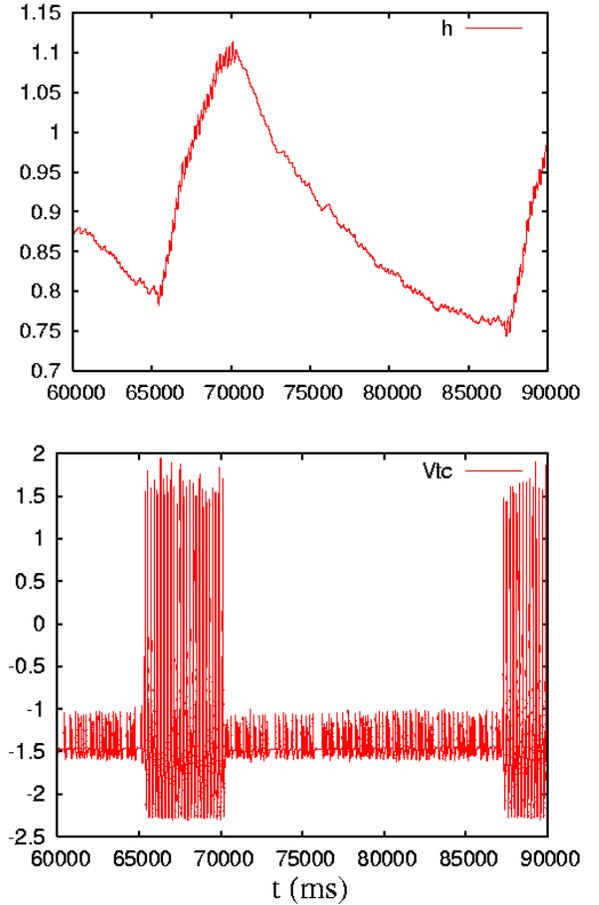}
\caption{\label{h_rev}(Color online) Top: Time course of $h$ during a spindle cycle. Bottom: The Membrane potential $v_{Tc}$  of the extended Hindmarsh-Rose model.}
\end{center}
\end{figure}
As Fig.\ \ref{h_rev} shows, the time course of the variable $h$ is very similar to the time course of  $s_{2}$ in the biophysical model.
To understand the influence of the variable $h$ we proceed in the same way as before. The timescale of the dynamics of $h$ 
is less than a tenth than the timescale of $z$. 
As declared above, the reason for the self-sustained oscillations is the after-hyperpolarization activation by the variable $z$. 
In the presence of $h$, the slowly varying parameter in the $(v,w)$ system is the sum $(z(t)+h(t))$. 
As $h(t)$ gets activated by $z(t)$, after some rebound bursts the sum $(z(t)+h(t))$ 
is below the threshold of the $(v,w)$ system, as a consequence, the oscillation terminates. 
When the excitor cell is inactive, there are no more inhibitory currents, that means that $t$ is inactive too, so $h$ decays slowly until it is small enough that the the system can get activated again by an input spike.

\subsection{Reciprocally coupled TC-RE neurons using the extended
Hindmarsh-Rose model}
As the $h$ current is absent in the biological RE neuron, we have to extend 
the Hindmarsh-Rose system by the $h$ equation (\ref{eq17d}) for the 
TC neuron only.
\begin{eqnarray}
\dot{v}_{TC}&=&w_{TC}-v_{TC}^3+3v_{TC}^2-z_{TC}-h_{TC}+I_{GABA}+I(t)\nonumber\\
\dot{w}_{TC}&=&1.8-5v_{TC}^2-w_{TC}\nonumber\\
\dot{z}_{TC}&=&0.006(4(v_{TC}+1.56)-z_{TC})\nonumber\\
\dot{h}_{TC}&=&-0.0004(h_{TC}+0.88(0.9-z))\nonumber\\[0.8ex]
\dot{v}_{RE}&=&w_{RE}-v_{RE}^3+3v_{RE}^2-z_{RE}+I_{GLU}\nonumber\\
\dot{w}_{RE}&=&1.8-5v_{RE}^2-w_{RE}\nonumber\\
\dot{z}_{RE}&=&0.006(4(v_{RE}+1.56)-z_{RE})
\label{eq16}
\end{eqnarray}
$I_{GABA}$ and $I_{GLU}$ are governed by  equations  (\ref{iga}) and (\ref{iglu}) in appendix \ref{secB1}, respectively.
The external input was modeled as described above for the biophysical model, 
with modified parameters adapted to the Hindmarsh-Rose model, 
$\bar{r}$ was $1/100$ and $\tau_{0}$ was $30ms$. 
If we stimulate our model by such a modified Poisson process, 
we have a computational model for the experiment of Le Masson {\itshape et al.} \cite{lem}, which allows us to compare our computational model with the experiment and the 
biophysical model \cite{des1}, so that we can validate our extension of the Hindmarsh-Rose model.

\begin{figure}[htbp]
\begin{center}
\includegraphics[angle=0,totalheight=10.5cm]{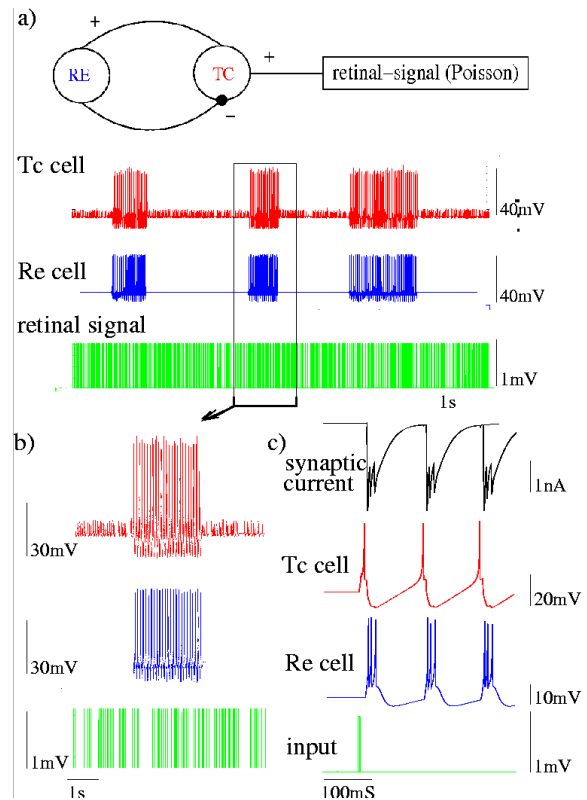}
\caption{\label{fig11}(Color online) Spontaneous spindle activity in our computational model 
(the numerical values of the voltages are scaled by a factor of 30 to match with the biophysical values). 
Similar rescalings have been also necessary in the underlying
Hindmarsh-Rose model (see, e.g.\ \cite{wil}).
{\bf a)} The computational circuit: A TC cell (excitor) modeled by the extended Hindmarsh-Rose model is reciprocally 
coupled to a model RE cell (inhibitor) represented by the original Hindmarsh-Rose model. 
The TC cell receives artificial synaptic retinal bombardment modeled by a Poisson distributed spike train. 
Like in the experiment the system shows spindle activity.
 {\bf b)} Detail of a). {\bf c)} Like in Fig.\ \protect\ref{fig8}.}
\end{center}
\end{figure}
A comparison of Fig.\  \ref{fig11} and 
(\cite{lem}, Fig.\ 2) 
shows that our system reproduces the experimental results of Le Masson {\itshape et al.} 
\cite{lem} quite well.
During the spindling state, the firing pattern of the TC cells (excitor), 
which is very different from the input, shows that the information transfer of the input is low. 
The question whether this low transfer is still reliable is calculated in the same way as for the biophysical system. 
\begin{figure}[hbtp]
\begin{center}
\includegraphics[angle=0,totalheight=6cm]{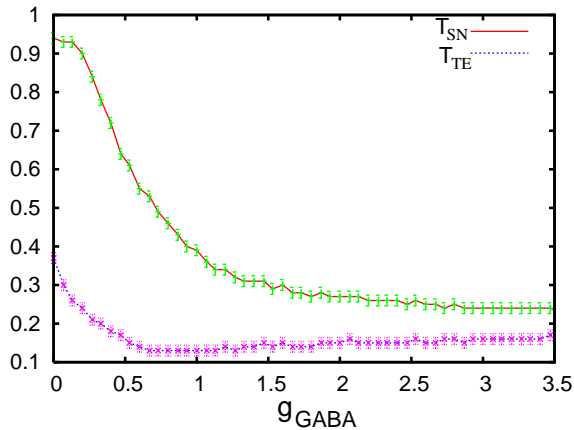}
\caption{\label{fig12}(Color online) Percentage of output spikes triggered by an input spike. 
An increase of the strength of the inhibitory coupling $g_{GABA}$ comes along with a 
significant decrease in the reliability of the information transfer. As in the experiment of 
\cite{lem} and the in the biophysical model, the transfer efficiency does not vary significantly. 
}
\end{center}
\end{figure}
With a strong inhibitory feedback, the  $T_{SN}$ was low, showing that most of the 
TC spikes where not triggered by an input spike and thus that the system is not transferring spikes in a 
one-to-one manner. This result encouraged us to screen the strength of inhibition to test if recurrent feedback 
inhibition could be a way to control the precision of input-output-transfer in a wide range, or if there is just 
a switching between a relay state and masquerading state. 
Our numerical investigation leads to the following results. An increase of the inhibitory coupling strength leads 
to a smooth decrease in $T_{SN}$ from a maximum value of around 1 to a minimum value of less than 0.3
(Fig.\ \ref{fig12}).
Despite this significant decrease in the reliability of spike transfer, the efficiency of spike transfer $T_{TE}$ was not significantly diminished by the strength of the 
inhibitory coupling (Fig.\ \ref{fig12}).  Thus inhibitory feedback has a direct decorrelating effect, 
which is able to reduce the reliability of the spike transfer.

\subsection{Dynamical behavior of coupled TC-RE neurons using the extended
Hindmarsh-Rose model}
As we argued in \ref{sec31} for the Hindmarsh-Rose model, Destexhe {\itshape et al.} \cite{des2} use singular approximation to characterize the spindle oscillations as a transition between a hyperpolarizing stationary phase and an oscillatory phase. Here we will apply this method to our simplified model of a thalamocortical oscillator. In Fig.\ \ref{fig11}, the $h$ current evolves according to a much slower time scale as the dynamics of the coupled TC-RE neurons. 

We assume $h$ to be a slow varying parameter in (\ref{eq16}) with $I(t)=0$, so we consider the limit $k_{2}\rightarrow 0$ in (\ref{eq1n}). We will consider $h$ as bifurcation parameter of (\ref{eq16}) in an interval between -1 and 1.5 as a parameter. In this range of $h$ the dynamical state of the system can be divided in three areas: For high values of h the system is in a stable resting state, for intermediate values of $h$ both  a stable resting state and a stable oscillatory state exist, separated by an unstable limit cycle. For negative values of $h$ the stable resting state gets unstable, while the stable oscillatory state persists. 
The transition between the stable fixed point and the oscillatory state occurs by a subcritical Hopf bifurcation \cite{hg2}. The existence of a region of parameters where stable solutions overlap is typical for a subcritical Hopf bifurcations; our system shows this typical behavior in a wide range of the bifurcation parameter.
\begin{figure}[hbtp]
\begin{center}
\includegraphics[angle=0,totalheight=6.5cm]{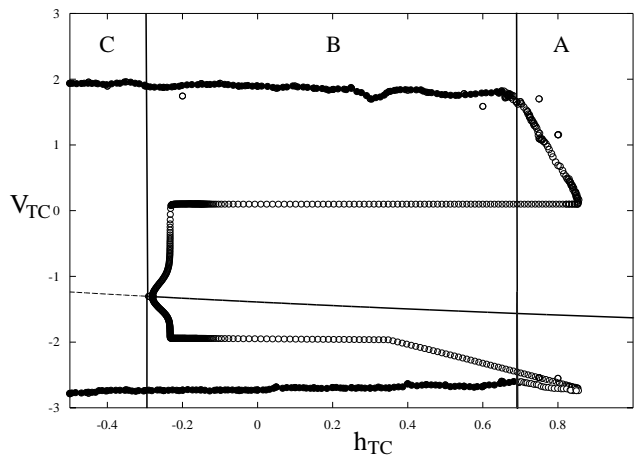}
\caption{\label{fig_bist}  Bifurcation diagram of (\ref{eq16}) with $h$ treated as a slowly varying parameter; solid lines indicate stable fixed points, dashed lines unstable fixed points, filled circles stable limit cycles, open circles unstable limit cycles. For extreme values of $h$, the system is monostable, i.e.\ either a stable fixed point or a stable limit cycle coexist. 
}
\end{center}
\end{figure}
In this states the system is bistable, i.e.\ a stable fixed point and a stable limit-cycle coexist (see Fig.\ \ref{fig_bist}). In this area, the state of the system depends on its history: If the initial conditions or the past state of the system lie within the basin of attraction of the limit cycle, the system exhibits stable oscillations; if the starting point lies within the basin of attraction of the fixed point, the system  rests on the fixed point. These two attractive areas are separated by an unstable limit cycle. Further, the system can be switched between these two states by an input pulse.

\subsection{Discussion of the transfer-properties of the extended Hindmarsh-Rose model}
The $h$ current switches the system between two existing dynamical states, namely a fixed point and a bistable state where a limit cycle and a fixed point coexist. Without this further current (i.e. $h=0$), two reciprocally coupled Hindmarsh-Rose neurons exhibit so-called hard self-excitation \cite{hak} as response to an input spike, resulting in stable self-sustained oscillations. The $h$ current makes this oscillatory state metastable, resulting in waxing and waning oscillations. An increase of the inhibition changes the responsivity of the system: Without, or with only little inhibition, the system responds to an input pulse with an output spike, so that spikes get transmitted in a one-to-one manner. With high gain in inhibitory feedback, the system responds to an input spike by an output burst (range B in Fig\ \ref{fig_bist}), followed by a silent period where transmission is totally blocked (range A in Fig\ \ref{fig_bist}). During the burst, the system is in an autonomous self-oscillatory state which masquerades the input, this leads to an increase of the $T_{SN}$. As the system is still excitable, but with a different responsivity the $T_{TE}$ does not change as strong as the $T_{SN}$. So the transmission behavior depends critically on the rate of rise and fall of the $h$ equation. This insight might also help to improve the biophysical model in terms of information transfer.

\section{Conclusions} 
We have shown that an extended Hindmarsh-Rose model is able to reproduce the 
behavior of a biological thalamocortical relay neuron
in the recent experiment of Le Masson {\sl et al.}
To gain more insight into the dynamical mechanisms,
a simplification or reduction of the
detailed biophysical model, which did not convincingly 
reproduce the experimentally observed 
decrease of the signal-to-noise ratio,
was necessary.
The widely used Hindmarsh-Rose model however does not show the
characteristic waxing and waning oscillations.
Further it does not exhibit the quiescence periods 
necessary for transfer of information.
Especially,  
we have analytically proven that the fast subsystem of the Hindmarsh-Rose model 
gets slaved by periodic forcing, which can lead to intrinsic bursting if the forcing is slow. 
To account for the waxing and waning mechanism,
we propose an extended Hindmarsh-Rose neuron model
for the TC cell, directly motivated from the 
biophysical model,
taking the low threshold calcium current
explicitely into account.

In this paper, we have demonstrated that our extended Hindmarsh-Rose model
serves as a
computational model for the setup in the
Le Masson {\sl et al}.\ experiment,
 showing that the information transfer can be adjusted within a wide range 
by the gain of the recurrent feedback inhibition. 
Our numerical investigations strongly suggest that the low threshold calcium current 
plays an important role for the information transfer in thalamocortical circuits.
From the technical simplicity of our approach and its agreement with the experiment,
this approach may give rise to similar models for neural systems where
a third timescale is apparent, and a second slow degree of freedom
is necessary to describe the dynamics.

\begin{acknowledgments}
This research has been supported by the 
Deutsche Forschungsgemeinschaft
(DFG)
within SFB 654 ``Plasticity and Sleep''.
The authors thank 
Jan Born, Lisa Marshall, and Matthias M\"olle 
for intensive and fruitful discussions. We thank an anonymous referee for detailed 
and constructive comments.
\end{acknowledgments}

\begin{appendix}
\section{Thalamocortical circuit}
\label{secA}
The thalamocortical circuit consists of a pair of 
TC 
and thalamic 
 RE 
neurons connected as shown in 
Fig \ref{fig1}. 
For each TC and RE cell several ion currents were included which will be described in detail here. 
All intrinsic currents are described by the same general equation \cite{hod}:
\begin{equation}
\label{eqA1}
I_{j}=\bar{g}_{j}m^{M}h^{N}(V-E_{j}).
\end{equation} 
Here the current $I_{j}$ is the product of the maximal conductance, 
$g_{j}$, activation $m$ and inactivation $h$ variable, 
and the difference between the membrane potential $V$ 
and the Nernst
reversal potential $E_{j}$. 
The powers $N$ and $M$ are the respective numbers
of ion channels to be open synchroneously. 
The gating of a membrane channel is described by a first order kinetic scheme
\begin{eqnarray}
\label{eqA2}
C &\xrightarrow{\alpha(V)}&  O\\
C &\xleftarrow{\beta(V)}  &  O,
\label{eqA3}
\end{eqnarray}  
where O and C stand for the open and the closed state of the gate, with $\alpha(V)$ and $\beta(V)$ as the transition rates. 
According to the Hodgkin-Huxley model \cite{hod}, the variables $m$ and $h$ represent the fraction of independent gates in the 
open state and are described by simple first order differential equations,
\begin{eqnarray}
\label{eqA4}
\dot{m}&=&-\frac{1}{\tau_{m}(V)}(m-m_{\infty}(V))\\
\dot{h}&=&-\frac{1}{\tau_{h}(V)}(h-h_{\infty}(V)).
\label{eqA5}
\end{eqnarray}
The steady states $m_{\infty}(V)$ and $h_{\infty}(V)$ and the time constants $\tau_{m/h}(V)$, respectively, can be written as functions of the 
transition rates $\alpha$ and $\beta$,
using $x\in\{m,h\}$, as
\begin{eqnarray}
\nonumber
\tau_{x}(V)&=&\frac{1}{\alpha_{x}(V)+\beta_{x}(V)}
\\
x_{\infty}(V)&=&\alpha_{x}(V)\tau_{x}(V).
\label{eqA6}
\end{eqnarray}

\subsection{The TC cell}\label{secA1}
Here we summarize the results of the previously mentioned investigations of TC cell membrane properties. The membrane potential of the 
TC neuron is given by
\begin{eqnarray}
\label{eqA7}
C_{m}\dot{V_{T}}&=&-I_{T_{L}}-I_{T}-I_{h}-I_{T_{Na}}-I_{T_{K}}\nonumber\\
                & &-I_{\rm GABA_{\rm a}}-I_{\rm GABA_{\rm b}}.
\end{eqnarray}
The area of a TC cell membrane is about $2.9*10^{-4}\;cm^{2}$ which is 
according to a cell capacity of $C_{m}=0.29nF$. 
All constants used in the simulations correspond to a cell of this size. 

\paragraph*{The leakage current $I_{T_{L}}$}
is the only passive current, and 
is governed by the Ohm law
\begin{eqnarray}
I_{T_{L}}=
g_{\rm T_{\rm L}}
(V_{T}-E_{L})
\label{eqA8}
\end{eqnarray} 
where $g_{\rm T_{\rm L}}=0.05mS$, $E_{L}=-86mV$.

All the other currents are active currents, and more complicated,
as detailed below.
\paragraph*{The sodium current $I_{Na}$}
 has the form
\begin{equation}
I_{T_{Na}}=g_{Na}m(t)^{3}h(t)(V_{T}-E_{Na}),
\label{eqA11}
\end{equation} 
with $g_{T_{Na}}=30mS$ and $E_{Na}=50mV$. In addition we have for the gating variables $m(t)$ and $h(t)$
\begin{eqnarray}
m_{\infty}(V_{T})&=&\frac{\alpha_{m}(V_{T})}{\alpha_{m}(V_{T})+\beta_{m}(V_{T})}\\\nonumber
h_{\infty}(V_{T})&=&\frac{\alpha_{h}(V_{T})}{\alpha_{h}(V_{T})+\beta_{h}(V_{T})}\\\nonumber
\tau_{m}(V_{T})&=&\frac{1}{\alpha_{m}(V_{T})+\beta_{m}(V_{T})}\\\nonumber
\tau_{h}(V_{T})&=&\frac{1}{\alpha_{h}(V_{T})+\beta_{h}(V_{T})},
\end{eqnarray}
where
\begin{eqnarray}
\alpha_{m}(V_{T})&=&0.32\frac{(V_{T}+37)}{1-\exp(-\frac{V_{T}+37}{4})}\\\nonumber
\beta_{m}(V_{T})&=&0.28\frac{(V_{T}+10)}{\exp(\frac{V_{T}+10}{5})-1}\\\nonumber
\alpha_{h}(V_{T})&=&0.128\exp(-\frac{V_{T}+33}{5})\\\nonumber
\beta_{h}(V_{T})&=&\frac{4}{\exp(-\frac{V_{T}+10}{5})+1}.
\end{eqnarray}
With equations (\ref{eqA4},\ref{eqA5}) 
the sodium current is completely described.
\paragraph*{The potassium current $I_{K}$}
 has the form
\begin{equation}
I_{T_{K}}=g_{T_{K}}m(t)^{4}(V_{T}-E_{k})
\end{equation}
with $g_{T_{K}}=2.mS$ and $E_{k}=-95mV$. The gating and transition variables are given by
\begin{eqnarray}
m_{\infty}(V_{T})&=&\frac{\alpha_{m}(V_{T})}{\alpha_{m}(V_{T})+\beta_{m}(V_{T})}\\\nonumber
\tau_{m}(V_{T})&=&\frac{1}{\alpha_{m}(V_{T})+\beta_{m}(V_{T})}\\\nonumber
\alpha_{m}(V_{T})&=&0.032\frac{(V_{T}+35)}{1-\exp(-\frac{V_{T}+35}{5})}\\\nonumber
\beta_{m}(V_{T})&=&0.5\exp(-\frac{V_{T}+40}{5}).
\end{eqnarray}
With 
 equation (\ref{eqA4})
 the potassium current is completely described.
\paragraph*{The low threshold $Ca^{2+}$ current $I_{T}$}
 is taken to be   
\begin{equation}
I_{T}=g_{Ca}m(t)^{3}h(t)(V-E_{T}),
\end{equation}
where $g_{Ca}=1.75mS$ and the reversal potential $E_{T}$ depends on the $Ca^{2+}$ concentration inside $([Ca]_{in})$ and outside $([Ca]_{out})$ the cell. It is defined by the 
Nernst equation
\begin{equation}
E_{T}=1000*\frac{RT}{2F}\ln\frac{[Ca]_{out}}{[Ca]_{in}},
\end{equation}  
where $R=8.31441\frac{J}{K \cdot mol}$, $T=309.15K$,$F=96489\frac{C}{mol}$, and $[Ca]_{out}=2mM$ is considered to be constant. In the remainder, all concentrations will be denoted by square brackets, 
following an usual convention (see e.g. \cite{rabino}).\\
The calcium dynamics in the cell are described by a simple model which was introduced in \cite{des3},
\begin{equation}
\dfrac{d[Ca]_{in}}{dt}=-AI_{T}-K_{T}[Ca]_{in}/([Ca]_{in}+K_{d}),
\label{eqA17}
\end{equation} 
where $A=0.179\frac{m{\rm mol}}{ms\cdot \mu A}, K_{T}=10^{-4}\frac{m {\rm mol}}{ms}$ and $K_{d}=K_{T}$.\\
The gating variables are governed by the following system which was proposed by Wang \cite{wan1}
\begin{eqnarray}
\dot{m}&=&({\tau_{m}(V_{T})})^{-1} \left( m-m_{\infty}(V_{T})\right)\nonumber\\
\dot{h}&=&\alpha_{1}(V_{T})\left( 1-h-d-K(V_{T})h\right)\nonumber\\
\dot{d}&=&\alpha_{2}(V_{T})\left( K(V_{T})(1-h-d)-d\right).
\end{eqnarray}
In addition, for this current the constants are 
\begin{eqnarray}
m_{\infty}(V_{T})&=&\big({1+\exp(\frac{V_{T}+65}{7.8})}\big)^{-1}\nonumber\\
\tau_{m}(V_{T})&=&0.15m_{\infty}(V_{T})\left(1.7+\exp(-\frac{V_{T}+30.8}{13.5})\right)\nonumber\\
\alpha_{1}(V_{T})&=&\exp\left(-\frac{(V_{T}+162.3)}{17.8} \right)/0.26\nonumber\\
K(V_{T})&=&\sqrt{0.25+\exp\left(\frac{(V_{T}+85.5)}{6.3}\right) }-0.5 \nonumber\\
\alpha_{2}(V_{T})&=&\frac{1}{\tau_{2}(V_{T})\left( K(V_{T})+1\right) }\nonumber\\
 \tau_{2}(V_{T})&=&\frac{62.4}{1+\exp\left(\frac{V_{T}+39.4}{30} \right) }.
\end{eqnarray}
The numerical values are fits to experimental data \cite{hug1}.

\paragraph*{The hyperpolarization-activated cation current $I_{h}$}
finally
is described by
\begin{equation}
I_{h}=g_{h}(S_{1}+S_{2})(F_{1}+F_{2})(V_{T}-E_{h}),
\end{equation}
where $g_{h}=0.15mS$ and $E_{h}=-43mV$. The gating variables for this current are governed by
\begin{eqnarray}
\dot{S_{1}}&=&\alpha_{s}(V_{t})S_{0}-\beta_{s}(V_{t})S_{1}+k_{2}\left(S_{2}-CS_{1} \right)\nonumber\\ 
\dot{F_{1}}&=&\alpha_{f}(V_{t})F_{0}-\beta_{f}(V_{t})F_{1}+k_{2}\left(F_{2}-CF_{1} \right) \nonumber\\
\dot{S_{2}}&=&-k_{2}\left( S_{2}-CS_{1} \right)\nonumber\\
\dot{F_{2}}&=&-k_{2}\left( F_{2}-CF_{1} \right),
\label{eqA21}
\end{eqnarray}
where $C=\frac{[Ca_{in}]}{5*10^{-4}}$, $k_{2}=4*10^{-4}$, the rate constants $\alpha_{s/f}(V_{t})$ and $\beta_{s/f}(V_{t})$ are related to the activation function $H_{\infty}$ and the time constants $\tau_{s/f}$
\begin{eqnarray}
\alpha_{s}(V_{t})&=&{H_{\infty}}/{\tau_{s}}\nonumber\\ 
\alpha_{f}(V_{t})&=&{H_{\infty}}/{\tau_{f}}\nonumber\\ 
\beta_{s}(V_{t}) &=&{(1-H_{\infty})}/{\tau_{s}}\nonumber\\ 
\beta_{f}(V_{t}) &=&{(1-H_{\infty})}/{\tau_{f}},
\end{eqnarray}
where
\begin{eqnarray}
H_{\infty}&=&\Big({1+\exp\Big(\frac{V_{t}+68.9}{6.5}\Big) }\Big)^{-1}\\ 
\tau_{s}  &=&\exp(({V_{t}+183.6})/{15.24})\nonumber\\ 
\tau_{f}  &=&{\exp\left(\frac{V_{t}+158.6}{11.2}\right)}
\Big/ \left({1+\exp\left(\frac{V_{t}+75}{5.5}\right)}\right). \nonumber
\end{eqnarray}
This makes the biophysical description of TC cell complete.
While it contains a lot of simplifications,
for example the single compartment assumption, it reproduces almost all important features of 
TC cells quite good. We will use this model in our computer experiment to verify theories obtained by using simplified models. 
For more details see \cite{des3}.
\\[-5ex]

\subsection{The RE cell}\label{secA2}
The membrane potential of the thalamic RE-neuron is governed by the cable equation:
\begin{eqnarray}
C_{m}\dot{V_{R}}&=&-I_{R_{L}}-I_{Ts}-I_{K[Ca]}-I_{CAN}-I_{R_{Na}}-I_{R_{K}}\nonumber\\
                & &-I_{GLU}.
\end{eqnarray}
For the RE cell the membrane capacity is $C_{m}=0.143nF$, the leakage conductance is 
$g_{R_{l}}=0.05mS$ and the reversal potential of the leakage current is $E_{L}=-80mV$. 
This model of thalamic RE cell was introduced by Destexhe  {\itshape et al.} in 
\cite{des4}.

\paragraph*{The equations for the sodium $I_{R_{Na}}$ and 
potassium $I_{R_{K}}$ current in the RE cell} 
are the same as for the TC cell, 
except that the conductance 
$g_{Na}=100mS$ 
is different.
The description of the RE cell potassium current is reached by putting 
$g_{K}=10mS$.

\paragraph*{The kinetics of the low threshold $Ca^{2+}$ current $I_{Ts}$}
 was
established as model for
 voltage clamp data on rat RE cells using a 
so-called 
$m^{2}h$ formalism \cite{hug1}
\begin{equation}
I_{Ts}=g_{Ts}m(t)^{2}h(t)(V_{R}-E_{Ts}),
\end{equation} 
where $g_{Ts}=1.75\mu S$ and $E_{Ts}$ depends on the calcium concentration in the same way as described for the TC cell. The calcium dynamics are described by 
equation (\ref{eqA11}), when $I_{T}$ is replaced by $I_{Ts}$. In addition we have for this current
\begin{eqnarray}
\\[-4ex]
m_{\infty}(V_{R})
\!&\!=\!&\!
\Big({1+\exp(-\frac{V_{R}+52}{7.4})}\Big)^{-1}
\nonumber\\\nonumber
\tau_{m}(V_{R})
\!&\!=\!&\!
1+ \frac{1}{3} \Big({\exp(\frac{V_{R}+27}{10})+\exp(-\frac{V_{R}+102}{15})}\Big)^{-1}\\\nonumber
h_{\infty}(V_{R})
\!&\!=\!&\!
\Big({1+\exp(\frac{V_{R}+80}{5})}\Big)^{-1}\\\nonumber
\tau_{h}(V_{R})
\!&\!=\!&\!
\frac{85}{3} 
+\frac{1}{3} \Big({\exp(\frac{V_{R}+48}{4})+\exp(-\frac{V_{R}+407}{50})}\Big)^{-1}.
\nonumber
\end{eqnarray}
The full kinetics of the $I_{T}$ current is reached if $m_{\infty}$, $\tau_{m}$, $h_{\infty}$ and $\tau_{h}$ are inserted in 
equations (\ref{eqA4},\ref{eqA5}).

\paragraph*{The calcium dependent currents:}
The RE cell possesses further two calcium dependent currents the slow $K^{+}$ current $I_{K[Ca]}$, and the slow nonspecific cation current $I_{CAN}$. According to \cite{des3} they are modeled as voltage-independent currents described by 
equation 
(\ref{eqA1}) with M=2 and N=0. The activation variable m obeys
\begin{equation}
\dot{m}=\alpha[Ca_{i}]^{2}(1-m)-\beta m,
\end{equation}
where $\alpha$ and $\beta$ are rate constants and $[Ca_{i}]$ is the intracellular calcium concentration. For $I_{K[Ca]}$ $\alpha=48 ms^{-1}mM^{-2}$ and $\beta=0.03ms^{-1}$, for  $I_{CAN}$ $\alpha=20 ms^{-1}mM^{-2}$ and $\beta=0.002ms^{-1}$.
With this the description of the RE cell is complete.
\\[-5ex]

\subsection{Synaptic currents}
\label{secA3}
Kinetic models of synaptic currents have been directly fitted to experimental data \cite{des1}, based on whole cell recorded synaptic currents obtained in hippocampal neurons \cite{oti}. The $GABA_{A}$ and $GLU$ (=glutamate) receptor mediated currents are both represented by a first order kinetic scheme \cite{des5}
\begin{equation}
C+T\leftrightarrows O
\end{equation} 
where the transition between closed (C) and open (O) states depend on the binding of the transmitter (T). The current is given by
\begin{equation}
I_{syn}=g_{syn}[O](t)(V(t)-E_{syn}),
\end{equation} 
where $g_{syn}$ is the maximal conductivity and $E_{syn}$ is the reversal potential. For AMPA receptors $E_{syn}=0mV$ and for $GABA_{A}$ receptors  $E_{syn}=-80mV$. $g_{syn}=0.05$ for $GABA_{A}$ and $g_{syn}=0.1$ for glutamate synapses. $[O](t)$ is the fraction of open channels
\begin{equation}
\dfrac{d[O](t)}{dt}=\alpha\left\lbrace 1-[O](t)\right\rbrace [T](t)-\beta[O](t),
\label{eqA30}
\end{equation} 
and $[T](t)$ is the concentration of transmitter released from time $t$ to time $t_{max}$,
\begin{equation}
[T](t)=A\Theta(t_{max}-t)\Theta(t),
\end{equation} 
where $\Theta (x)$ is the Heaviside function. The synaptic parameter values used in the simulation are 
$A=0.5$ and $t_{max}=3ms$ for $GABA_{A}$ and $9ms$ for $AMPA$ synapses, the rate constants were chosen as $\alpha=5ms$ and $\beta=0.16ms$ for $GABA_{A}$ synapses and $\alpha=0.94ms$ and $\beta=0.18ms$ for $GLU$ synapses.
The activation scheme of $GABA_{B}$ receptors is more complex, as it involves the activation of $K^{+}$ channels by $G$ proteins (for details see \cite{dut}). The model used by Le Masson {\itshape et al.} \cite{lem} is a modified version of the $GABA_{B}$ kinetic model introduced in \cite{des6}
\begin{eqnarray}
R_{0}+T   &\rightleftarrows& R \leftrightarrows D\\
  R+G_{0} &\leftrightarrows& RG\rightarrow R+G \\
       G  &\rightarrow     & G_{0}\\
C_{1}+4G  &\leftrightarrows& O,
\end{eqnarray} 
where the transmitter, $T$, binds to the receptor, $R_{0}$, leading to its activated form, $R$, and desensitized form, $D$. The $G$ protein is transformed from an inactive (GDP-bound) form, $G_{0}$ to an activated form, G, catalyzed by $R$. Finally, $G$ binds to open the $K^{+}$ channel, with four independent binding sites. With some assumptions (see \cite{des1}) the kinetic model for this system reduces to
\begin{eqnarray}
I_{GABA_{B}}   &=&g_{GABA_{B}}\frac{[G]^{4}}{[G]^{4}+K_{d}}\left[ V(t)-E_{k}\right] \\\nonumber
\frac{d[R]}{dt}&=&K_{1}\left( 1-[R]\right) [T]-K_{2}[R]\\\nonumber
\frac{d[G]}{dt}&=&K_{3}[R]-K_{4}[G],
\end{eqnarray} 
where $[R]$ is the fraction of activated receptors and $[G]$ is the concentration of $G$ proteins. In these equations $g_{GABA_{B}},\;K_{1}=0.52,\;K_{2}=0.0013,\;K_{3},\;K_{4}=0.033$ and $K_{d}=100.$\\
The strengths of the synaptic couplings $g_{GABA_{A/B}}$ and $g_{AMPA}$ are varied in different simulations.
\\[-5ex]

\section{The reduced model}
\label{secB}

\subsection{Synaptic Currents:}
\label{secB1}
The simplest way to model synaptic currents is a two state scheme of the binding of a neurotransmitter $T$ to postsynaptic receptors \cite{des1}. So the receptor mediated currents are both represented by a first order kinetic scheme
\begin{displaymath}
C+T \leftrightarrows O,
\end{displaymath}
where the transition between closed ($C$) and open ($O$) states depend on the binding of the transmitter ($T$) with forward and backward rates $\alpha$ and $\beta$ respectively. As in \cite{des7} we only consider  voltage-independent rate constants. With this assumptions we get the following kinetic equation
\begin{equation}
\frac{d[O]}{dt}=\alpha(1-[O])Tr(t)-\beta [O]=\alpha Tr(t)-[O] \alpha Tr(t)-\beta [O].
\end{equation}
The synaptic current is given by
\begin{equation}
I_{syn}=g_{syn}[O](t)(V(t)-E_{syn}),
\end{equation}
where $g_{syn}$ is the maximal conductivity and $E_{syn}$ is the reversal potential. For glutamate receptors $E_{syn}=0mV$ and for $GABA_{A}$ receptors  $E_{syn}=-50mV$. 
The release and clearance of transmitter are extremely fast processes compared to the open/close kinetics, resulting in a very brief presence of transmitter in the synaptic cleft \cite{cle1}, which allows us to consider $O$ to be constant during the transmitter release. For simplicity we assume the transmitter release time course to be a square pulse $A\Theta(V_{pre})$ which occurs when the presynaptic neuron fires a spike, i.e. when the presynaptic potential $V_{pre}$ gets depolarized towards positive potentials.
If we assume that initially all synaptic channels are in the closed state, we get
\begin{equation}
\frac{d[O]}{dt}=\gamma\alpha A\Theta(V_{pre})-\beta [O],
\end{equation}
were $\gamma=\alpha A$.
As $GABA_{B}$ synapses contribute less than a tenth of the total inhibitory synaptic current, they are neglected here.
The rate constants and amplitude were taken from the biophysical model \ref{secA3}. The constants for the $GABA$ synapse were modified because of the absence of the $GABA_{B}$ synapse: $\alpha=5ms^{-1}$ and $\beta=0.05ms^{-1}$. For the $GLU$ synapse we used the same constants as in the biophysical model: $\alpha=0.94ms^{-1}$ and $\beta=0.18ms^{-1}$ . We chose  $A=0.5$, and respectively $\gamma=2.5$ for $GABA$ synapses and $\gamma=0.47$ for glutamate synapses. The synaptic currents are governed by the following equations:
\begin{eqnarray}
I_{GABA}=g_{GABA}[O]_{GABA}(v_{TC}-2.5) \label{iga}\\
I_{GLU}=g_{GLU}[O]_{GLU}v_{RE},\label{iglu}
\end{eqnarray}
the reversal potential of the $GABA_{A}$ synapse was rescaled by a factor of 30 to adapt it to the scale of the Hindmarsh-Rose model.  
\subsection{Transformation to a Lienard system}
\label{secB2}
\noindent
We begin with the autonomous $(v,w)$ subsystem 
(\ref{eq7},\ref{eq8}),
\begin{eqnarray}
\dot{v}(t)&=&w-v^3+3v^2\label{eqB1}\\
\dot{w}(t)&=&1.8-5v^2-w\label{eqB2}.
\end{eqnarray}
Then
$\ddot{v}=\dot{w}+(6v-3v^{2})\dot{v}$,
$\dot{w}=\ddot{v}-(6v-3v^{2})\dot{v}$,
and
\begin{eqnarray} w=-\ddot{v}+(6v-3v^{2})\dot{v}+1.8-5v.
\label{eqB3}
\end{eqnarray}
Inserting
(\ref{eqB3}) in (\ref{eqB1}) 
yields
\begin{eqnarray}
\label{eqB4}
0&=&\ddot{v}+(1.8-6v+3v^{2})\dot{v}+(v^{3}+2v^{2}-1.8)\nonumber\\
\Leftrightarrow 0&=&\ddot{v}+f(v)\dot{v}+g(v).
\end{eqnarray}
(\ref{eqB4}) is the so called Lienard form of 
equations 
(\ref{eqB1})  and 
(\ref{eqB2}), 
where $f(v)=1-6v+3v^{2}$ and $g(v)=v^{3}+2v^{2}-1.8$.
\\[-5ex]

\subsection{Stability of the equilibrium points}
\label{secB3}
The stability of equilibrium points in the fast $(v,w)$-subsystem is investigated by  a linear approximation to the system 
(\ref{eq7},\ref{eq8}).
Suppose the equilibrium point has the $v$-coordinate $v_{0}$, then the linear approximation is:
\begin{equation}
\left( \begin{array}{llll} \dot{x} \\
                           \dot{y} \end{array} \right) 
=A(v_{0}) \left( \begin{array}{llll} x \\
y \end{array} \right),  
\end{equation}
where $(x,y)$ are new coordinates whose origin is in the equilibrium point, and $A(v_{0})$ is the Jacobian in $v_{0}$
\begin{equation}
A(v_{0})=\left( \begin{array}{cccc} -3v_{0}^{2}+6v_{0} &  1 \\
                                    -10v_{0}           & -1  \end{array} \right).
\end{equation} 
The kind of equilibrium point may be determined by the signs of the 
trace 
${\rm Tr}(A(v_{0})) =-3v_{0}^{2}+6v_{0}-1$
and determinant
${\rm Det}(A(v_{0}))=3v_{0}^{2}+4v_{0}$
\cite{hind,arg}.
The eigenvalues $\lambda_{1/2}$ of the Jacobian $A(v_{0})$ are expressed by
\\[-3ex]
\begin{equation}
\lambda_{1/2}=\frac{1}{2}\!\left\lbrace {\rm Tr}(A(v_{0}))\pm\sqrt{({\rm Tr}(A(v_{0})))^{2}
\!-\! 4{\rm Det}(A(v_{0}))} \right\rbrace 
\end{equation}

Table \ref{tab1} gives the type of equilibrium point according to the region $v_{0}$ belongs to;
here ${\rm Tr}_{-}=(3-\sqrt{6})/3$ and ${\rm Tr}_{+}=(3+\sqrt{6})/3$ 
are the negative and postive zeroes of ${\rm Tr}(A(v_{0}))$, respectively.
\begin{table}[htbp]
\caption{\label{tab1}Equilibrium points depending on $v_0$ (see text). 
Only stable nodes and unstable spirals and saddle nodes occur.}
\begin{tabular}{ccccc}
\hline\hline
region & values of $v_{0}$  & sign of              &  sign of             &
 type of 
  \\
       &                    & 
${\rm Tr}A(v_{0})$ & ${\rm Det}A(v_{0})$&node     
\\
\hline\hline
 I     & $v_{0}<-4/3$ & $-$ & $+$ & stable  \\ 
 II    & $-4/3<v_{0}<{\rm Tr}_{-}$ & $-$ & $-$ & saddle \\ 
 III    & ${\rm Tr}_{-}<v_{0}<0$& $+$ & $-$ &  saddle \\ 
 IV   & $0<v_{0}<{\rm Tr}_{+}$ & $+$ & $+$ &  unstable spiral \\ 
  V    & ${\rm Tr}_{+}<v_{0}$   & $+$ & $-$ &  saddle \\
\hline\hline
\end{tabular} 
\end{table}
\\[-2ex]

\subsection{Periodic solutions for periodic forcing\label{secB4}}
The conditions for constants $b,m,M>0$ are
\begin{enumerate}[(i)]
\item{for $|v|\ge b,f(v)>m$}
\item{$ \forall\; v \in \mathbb{R}, f(v)>-M$}
\item{for $|v|\ge b, vg(v)>0$}
\item{$g(v)$ is monotone increasing in $(-\infty,-b)$}
\item{$|g(v)|\to \infty$ for $|v|\to \infty$}
\item{$g(v)/G(v) \to 0$ for $|v| \to \infty$, \\ where $ G(v) =\int_{0}^{v} g(u)du$. }
\end{enumerate}
The proof can be sketched as follows:
\\[-4ex]
\begin{enumerate}[(a)]
\item{The existence of $M$  is easy to establish as f(v) has a minimum at $v=1$, further as $f(v)$ is a convex function this minimum is the global one. If a $M> \vert f(1)=-2 \vert$ is chosen (ii) holds.} 
\\[-5ex]
\item{Set $M=m$, then $b_{1}$ may be chosen arbitrarily.} 
\\[-5ex]
\item{Next $vg(v)$ is a fourth degree polynomial where the coefficient of the $v^{4}$-term is positive. So there exists a $b_{2}$ with $vg(v)>0$ if $v>b_{2}>0$.}
\\[-5ex]
\item{As $g(v)$ is a third degree polynomial, there is a $b_{3}$, such that $g(v)$ is monotonously increasing in $(-\infty,-b_{3})$ and $(b_{3},\infty)$.}
\\[-5ex]
\end{enumerate}
Choose $b={\rm max}(b_{1},b_{2},b_{3})$ then (i), (iii) and (iv) hold. 
Finally (v) and (vi) are obvious; so the proof is complete.
\end{appendix}


\begin{thebibliography}{10}
\bibitem{sws} L.Marshall, M. M\"olle, and J. Born, \JTIT{Spindle and slow wave rhythms at slow wave sleep transitions are linked to strong shifts in the cortical direct current potential,} Neuroscience {\bf 121}, 1047-1053 (2003).
\bibitem{lem} G. Le Masson, S. Renaud-Le Masson, D. Debay, T. Bal,  \JTIT{Feedback inhibition controls spike transfer in hybrid thalamic circuits,} Nature (London) {\bf 417}, 854, (2002).
\bibitem{des1} A. Destexhe, T. Bal, D. A. McCormick, T. J. Sejnowski, \JTIT{Ionic Mechanisms Underlying Synchronized Oscillations and Propagating Waves in a Model of Ferret Thalamic Slices,} J. Neurophys. {\bf 76} 2049, (1996).
\bibitem{des2} A. Destexhe, D. A. McCormick, T. J. Sejnowski, \JTIT{A Model for 8-10 Hz Spindling in Interconnected Thalamic Relay and Reticularis Neurons,} Biophysical Journal {\bf 65} 2473, (1993).
\bibitem{amo} A. Amon, M. Nizette, M. Lefranc, T. Erneux, \JTIT{Bursting oscillations in optical parametric oscillators,} Phys. Rev. A {\bf 68}, 23801, (2003). 
\bibitem{str}R. Straube, S. C. M\"uller, M.J.B. Hauser, \JTIT{Bursting Oscillations in the Revised Mechanism of the Hemin - Hydrogen Peroxide - Sulfite Oscillator,} Zeitschrift f\"ur Physikalische Chemie {\bf 217}, 1427, (2003).
\bibitem{sri} S. Raghavachari, J. A. Glazier, \JTIT{Waves in Diffusively Coupled Bursting cells,} Phys. Rev. Lett {\bf 82}, 2991 (1999).
\bibitem{dipl} J\"org Mayer, Modellierung und Analyse neuronaler Dynamiken im Thalamus, Diploma thesis,  Kiel, 2004.
\bibitem{izh} Eugene M. Izhikevich, \JTIT{Neural Excitability Spiking and Bursting,} International Journal of Bifurcation and Chaos {\bf 10}, 1171 (2000).
\bibitem{sher} Sherman S. M., Guillery R. W., \JTIT{Functional organisation of thalamocortical relays,} J. Neurophys. {\bf 76}, 1367 (1996).
\bibitem{pin} J. Pinell, Biopsychologie, Spektrum Akademischer Verlag, 4th edition, 1997.
\bibitem{hg1} H.G.Schuster, Complex Adaptive Systems, Scator, Saarbr\"ucken 2001.
\bibitem{hg2} H.G.Schuster, W. Just, Deterministic Chaos, 4th edition, Wiley-VCH, 2005.
\bibitem{hind} J.L. Hindmarsh, R.M. Rose, \JTIT{A model of neuronal bursting using three coupled first order differential equations,} Proc. R. Soc. Lond. B {\bf 221}, 87 (1984).
\bibitem{rabino} M. Rabinovich, R. Huerta, M. Bazhenov, A. K. Kozlov, and H. D. I. Abarbanel, \JTIT{Computer simulations of stimulus dependent state switching in basic circuits of bursting neurons,} Phys. Rev. E {\bf 58}, 6418-6430 (1998).
\bibitem{bal1} T. Bal, M. von Krosigk,D. A. Mc Cormick, \JTIT{Synaptic and membrane mechanisms underlying synchronized oscillations in the ferret LGNd in vitro,} J. Physiol. Lond {\bf 483}, 641 (1995).
\bibitem{des3} A. Destexhe, A. Babloyantz, T. J. Sejnowski, \JTIT{Ionic Mechanisms for Intrinsic Slow Oscillations in Thalamic Relay Neurons,} Biophysical Journal {\bf 65}, 1538 (1993).
\bibitem{hug1} J. R. Huguenard, D. A. Prince, T. J. Sejnowski, \JTIT{A novel T-type current underlies prolonged $Ca^{2+}$-dependent bursts firing in GABAergic neuroons of rat thalamic reticular nucleus,} Journal of Neuroscience {\bf 12}, 3804 (1994).
\bibitem{bal2} T. Bal, D. A. McCormick, \JTIT{Mechanisms of oscillatory activity in guinea-pig nucleuss reticularis in vitro: a mammalian pacemaker,} J. Physiol. Lond {\bf 468}, 669 (1993).
\bibitem{hod} A. L. Hodgkin, A. F. Huxley, \JTIT{A quantitative description of membrane current and its application to conduction and excitation in nerve,} J. Physiol. (London) {\bf 117}, 500 (1952).
\bibitem{dham} Mukeshwar Dhamala, Viktor K. Jirsa, Mingzhou Ding, \JTIT{Transitions to Synchrony in Coupled Bursting Neurons,} Phys. Rev. Lett. {\bf 92}, 028101 (2004).
\bibitem{pon}L. S. Pontyagin, \JTIT{Asymptotic behaviour of the solutions of systems of differential equations with a small parameter in the higher derivatives,} Am. Math. Soc. Trans. Sec. II {\bf 18}, 275 (1967).
\bibitem{zee} E. C. Zeeman, \JTIT{Differential equations for the heart beat and nerve impulse,} {\itshape In} Dynamical Systems, M. Peixoto, editor, Academic Press New York 683 (1973).
\bibitem{wil} H. R. Wilson, \JTIT{Simplified Dynamics of Human and Mamalian Neocortical Neurons,} Journal of Theoretical Biology {\bf 200}, 375 (1999).
\bibitem{auto} E. Doedel, Cong. Num. {\bf 30}, 265 (1981).
\bibitem{xpp} B. Ermentrout,  Simulating, Analyzing, and Animating Dynamical Systems: A Guide to XPPAUT for Researchers and Students, SIAM Books,  Soc. for Industrial \&  Applied Math,  1st edition,  2002.  
\bibitem{wang} X.-J. Wang, \JTIT{Genesis of bursting oscillations in the Hindmarsh-Rose model and homoclinicity to a chaotic saddle,} Physica D {\bf 62}, 263 (1993).
\bibitem{tra} R. D. Traub, R. Miles, Neuronal Networks of the Hippocampus, Cambridge University Press, 1991.
\bibitem{far}  Miklos Farkas, Periodic Motions, Applied Mathematical Sciences {\bf 104}, 171 (1994).
\bibitem{des7} A. Destexhe, M. Rudolph, \JTIT{Extracting Information from the Power Spectrum of Synaptic Noise,} Journal of Computational Neuroscience  {\bf 17}, 327 (2004).
\bibitem{cle1} J. D. Clements, \JTIT{Transmitter time course in the synaptic cleft: Its role int central synaptic function,} Trends Neurosci. {\bf 19}, 163 (1996)
\bibitem{ros}  Michael G. Rosenblum, Arkady S. Pikovsky, \JTIT{Controlling Synchronization in an Ensemble of Globally Coupled Oscillators,} Phys. Rev. Lett. {\bf 92}, 114102 (2004).
\bibitem{hak} H. Haken, Synergetics, An Introduction, Springer, 1977.
\bibitem{ahl} Ahlsen G., Lindstr\"om S., \JTIT{Interaction between inhibitory pathways to principal cells in the lateral geniculate nucleus of the cat,} Exp. Brain. Res. {\bf 58}, 134 (1985).
\bibitem{wan1} X. J. Wang, J. Rinzel, M. A. Rogawski, \JTIT{A Model of the T-type calcium current and the low threshold spike in thalamic neurons,} J. Neurophysiol. {\bf 66}, 850  (1991).
\bibitem{des4} A. Destexhe, A. Babloyantz,  \JTIT{A model of the inward current $I_{h}$ and its possible role in thalamocortical oscillations,} Neuro Report {\bf 4}, 223 (1993).
\bibitem{oti} T. Otis, I. Mody, \JTIT{Modulation of decay kinetics and frequency of $GABA_{A}$ receptor-mediated spontaneous inhibitory postsynaptic currents in hippocampal neurons,} Neuroscience {\bf 49}, 13 (1992).
\bibitem{des5} A. Destexhe, Z. F. Mainen, T. J. Sejnowski, \JTIT{Synthesis of Models for Excitable Membranes, Synaptic Transmission and Neuromodulation using a Common Kinetic Formalism,} J. of Comp. Neuroscience {\bf 1}, 195 (1994). 
\bibitem{dut} P. Dutar, R. A. Nicoll, \JTIT{A physiological role for $ GABA_{B}$ receptors in the central nervous system,}  Nature (London) {\bf 332}, 158 (1988).
\bibitem{des6} A. Destexhe, T. J. Sejnowski, \JTIT{G-protein activation kinetics and spill over of GABA may account for differences between responses in the hippocampus and thalamus,} Proc. Nat. Acad. Sci. USA {\bf 92}, 9515 (1995).  
\bibitem{arg} J.\ H.\ Argyris, G.\ Faust, M.\ Haase, Exploration of Chaos: An Introduction for Natural Scientists and Engineers, Elsevier, Amsterdam, 1994. 

\end{thebibliography}
\end{document}